\documentstyle[12pt,aps]{revtex}
\newcommand{\bA}{\mbox{\boldmath $A$\unboldmath}}
\newcommand{\bn}{\mbox{\boldmath $n$\unboldmath}}
\newcommand{\bl}{\mbox{\boldmath $l$\unboldmath}}
\newcommand{\bm}{\mbox{\boldmath $m$\unboldmath}}
\newcommand{\cD}{\cal D}
\newcommand{\cL}{\cal L}
\newcommand{\cLd}{{\cal L}^{\dagger}}
\newcommand{\tkr}{2\kappa (r-r_H)}
\newcommand{\sta}{\sin\theta}
\newcommand{\cta}{\cos\theta}
\newcommand{\sda}{\sin^2\theta}

\newcommand{\coa}{\cot\theta}
\newcommand{\sqd}{\sqrt{2}}

\newcommand{\pr}{\frac{\partial}{\partial r}}
\newcommand{\pv}{\frac{\partial}{\partial v}} 
\newcommand{\pta}{\frac{\partial}{\partial \theta}}
\newcommand{\pvi}{\frac{\partial}{\partial \varphi}}
\newcommand{\pdr}{\frac{\partial^2}{\partial r^2}}
\newcommand{\pdta}{\frac{\partial^2}{\partial \theta^2}}
\newcommand{\pdvi}{\frac{\partial^2}{\partial \varphi^2}}
\newcommand{\pdvr}{\frac{\partial^2}{\partial v \partial r}}

\newcommand{\spr}{\frac{\partial}{\partial r_*}}
\newcommand{\spv}{\frac{\partial}{\partial v_*}}
\newcommand{\spdr}{\frac{\partial^2}{\partial r_*^2}}
\newcommand{\spdvr}{\frac{\partial^2}{\partial r_* \partial v_*}}

\begin{document}
\pagestyle{myheadings}
\markboth{}{Wu and Cai}
\draft

\title{\Large {\bf Asymmetry of Hawking Radiation of Dirac Particles \\
in a Charged Vaidya - de Sitter Black Hole}}
\author{S. Q. Wu\thanks{E-mail: sqwu@iopp.ccnu.edu.cn} and 
X. Cai\thanks{E-mail: xcai@ccnu.edu.cn}}
\address{Institute of Particle Physics, Hua-Zhong 
Normal University, Wuhan 430079, China}
\date{\today}
\maketitle

\vskip 1cm
\begin{center}{\bf ABSTRACT}\end{center}
\begin{abstract}
\widetext
The Hawking radiation of Dirac particles in a charged Vaidya - de Sitter
black hole is investigated by using the method of generalized tortoise 
coordinate transformation. It is shown that the Hawking radiation of Dirac 
particles does not exist for $P_1, Q_2$ components, but for $P_2, Q_1$ 
components it does. Both the location and the temperature of the event 
horizon change with time. The thermal radiation spectrum of Dirac particles 
is the same as that of Klein-Gordon particles.

{\bf Key words}: Hawking effect, Dirac equation, evaporating black hole 
\end{abstract} 
\pacs{PACS numbers: 97.60.Lf, 04.70.Dy}

\newpage
\baselineskip 20pt

\section{Introduction}

Hawking's investigation of quantum effects \cite{Hawk}interpreted as the
emission of a thermal spectrum of particles by a black hole event horizon
sets a significant landmark on black hole physics. In the last few decades, 
much work has been done in the Hawking effect of black holes in different 
types of space-time, such as Vaidya \cite{KCKY}, Kerr-Newman \cite{WC} and
NUT-Kerr-Newman-de Sitter \cite{AM} space-times. The thermal radiation of 
Dirac particles, especially with the aid of Newman-Penrose formalism 
\cite{NP}, has been also investigated in some spherically symmetric and 
non-static black holes \cite{Zhel}. However, most of these studies 
concentrated on the spin state $p=1/2$ of the four-component Dirac spinors. 
Recently, the Hawking radiation of Dirac particles of spin state $p= -1/2$ 
attracts a little more attention \cite{LMZ}.

In this paper, we investigate the Hawking effect of Dirac particles in the 
Vaidya-type black hole by means of the generalized tortoise transformation 
method. We consider simultaneously the limiting forms of the first order 
form and the second order form of Dirac equation near the event horizon 
because the Dirac spinors should satisfy both of them. From the former, we 
can obtain the event horizon equation, while from the latter, we can derive 
the Hawking temperature and the thermal radiation spectrum of electrons.
Our results are in accord with others. With our new method, we can prove 
rigorously that the Hawking radiation does not exist for $P_1, Q_2$ 
components of Dirac spinors. The origin of this asymmetry of the Hawking 
radiation of different spinorial components probably stem from the asymmetry 
of space-time in the advanced Eddington-Finkelstein coordinate system. As a 
by-product, we point out that there could not have been any new quantum thermal 
effect \cite{LMZ} in the Hawking radiation of Dirac particles in any 
spherically symmetric black hole whether it is static or non-static. 

The paper is organized as follows: In section 2, we work out the spinor form 
of Dirac equation in the Vaidya-type black hole, then, we obtain the event
horizon equation in Sec. 3. The Hawking temperature and the thermal radiation 
spectrum are derived in Sec. 4 and 5, respectively. Sec. 6 is devoted to 
some discussions.

\section{Dirac equation}

The metric of a charged Vaidya - de Sitter black hole with the cosmological 
constant $\Lambda$ is given in the advanced Eddington-Finkelstein coordinate 
system by 
\begin{equation}
ds^2 = 2dv(G dv -dr) -r^2(d\theta^2 +\sin^2\theta d\varphi^2) \, , 
\end{equation}
and the electro-magnetic one-form is
\begin{equation}
\bA=\frac{Q}{r}dv \, 
\end{equation}
where $2G = 1 -\frac{2M}{r} +\frac{Q^2}{r^2} -\frac{\Lambda}{3}r^4$, in which 
both mass $M(v)$ and electric charge $Q(v)$ of the hole are functions of 
the advanced time $v$.

We choose such a complex null-tetrad $\{\bl, \bn, \bm, \overline{\bm}\}$  
that satisfies the orthogonal conditions $\bl \cdot \bn = -\bm \cdot 
\overline{\bm} = 1$. Thus the covariant one-forms can be written as
\begin{equation}
\begin{array}{ll}
\bl = dv \, , ~~
&\bm = \frac{-r}{\sqd}\left(d\theta +i\sta d\varphi\right) \, , \\
\bn = G dv -dr \, , ~~ 
&\overline{\bm} = \frac{-r}{\sqd}\left(d\theta -i\sta d\varphi\right) \, . 
\end{array}
\end{equation}
and their corresponding directional derivatives are
\begin{equation}
\begin{array}{ll}
D = -\pr \, ,  ~~
&\delta = \frac{1}{\sqd r}\left(\pta +\frac{i}{\sta}\pvi\right) \, , \\ 
\Delta = \pv +G\pr \, , ~~
&\overline{\delta} = \frac{1}{\sqd r}\left(\pta -\frac{i}{\sta}\pvi\right) \, .  
\end{array}
\end{equation}

It is not difficult to determine the non-vanishing Newman-Penrose 
complex coefficients \cite{NP} in the above null-tetrad 
\begin{equation}
\rho = \frac{1}{r} \, , ~~\mu = \frac{G}{r} \, , 
~~\gamma = -\frac{G_{,r}}{2} \, ,
~~\beta = -\alpha = \frac{\coa}{2\sqd r}  \, .  
\end{equation}

Inserting for the following relations among the Newman-Penrose  
spin-coefficients \footnote{Here and hereafter, we denote $G_{,r} 
= dG/dr$, etc.} 
\begin{equation}
\begin{array}{ll}
\epsilon -\rho = -\frac{1}{r} \, , ~~
&\tilde{\pi} -\alpha = \frac{\coa}{2\sqd r} \, , \\
\mu -\gamma = \frac{G}{r} +\frac{G_{,r}}{2} \, , ~~
&\beta -\tau = \frac{\coa}{2\sqd r} \, ,
\end{array}
\end{equation}
and the electro-magnetic potential
\begin{equation}
\bA \cdot \bl = 0 \, ,~~\bA \cdot \bn = Q/r \, ,
~~\bA \cdot \bm = -\bA \cdot \overline{\bm} = 0 \, ,
\end{equation}
into the spinor form of the coupled Chandrasekhar-Dirac equation \cite{CP}, 
which describes the dynamic behavior of spin-$1/2$ particles, namely
\begin{equation}
\begin{array}{ll}
&(D +\epsilon -\rho +iq\bA\cdot\bl)F_1 +(\overline{\delta} +\tilde{\pi} 
-\alpha  +iq\bA\cdot\overline{\bm})F_2 =\frac{i\mu_0}{\sqd}G_1 \, , \\
&(\Delta +\mu -\gamma  +iq\bA\cdot\bn)F_2 +(\delta +\beta -\tau 
+iq\bA\cdot\bm)F_1 =\frac{i\mu_0}{\sqd}G_2 \, ,\\
&(D +\epsilon^* -\rho^* +iq\bA\cdot\bl)G_2 -(\delta +\tilde{\pi}^*
-\alpha^* +iq\bA\cdot\bm)G_1 =\frac{i\mu_0}{\sqd}F_2 \, , \\
&(\Delta +\mu^* -\gamma^* +iq\bA\cdot\bn)G_1 -(\overline{\delta}
+\beta^* -\tau^* +iq\bA\cdot\overline{\bm})G_2 =\frac{i\mu_0}{\sqd}F_1 \, ,
\end{array}
\end{equation}
where $\mu_0$ and $q$ is the mass and charge of Dirac particles, one obtains 
\begin{equation}
\begin{array}{ll}
-\left(\pr + \frac{1}{r}\right)F_1 +\frac{1}{\sqd r} {\cL} F_2 
= \frac{i\mu_0}{\sqd} G_1 \, , 
&\frac{1}{2r^2} {\cD} F_2  +\frac{1}{\sqd r} {\cLd} F_1 
= \frac{i\mu_0}{\sqd} G_2 \, , \\
-\left(\pr + \frac{1}{r}\right)G_2 -\frac{1}{\sqd r} {\cLd} G_1 
= \frac{i\mu_0}{\sqd} F_2 \, , 
&\frac{1}{2r^2} {\cD} G_1 -\frac{1}{\sqd r} {\cL} G_2 
= \frac{i\mu_0}{\sqd} F_1 \, , \label{DCP}
\end{array}
\end{equation}
in which we have defined operators
$${\cD} = 2r^2\left(\pv +G\pr \right) +(r^2G)_{,r} +2iqQr  \, ,$$
$${\cL} = \pta +\frac{1}{2}\coa -\frac{i}{\sta}\pvi  \, , 
~~{\cLd} = \pta +\frac{1}{2}\coa +\frac{i}{\sta}\pvi  \, .$$

One can observe that the Chandrasekhar-Dirac equation (8) could be 
satisfied by identifying $Q_1$, $Q_2$, $qQ$ with $P_2^*$, $-P_1^*$, $-qQ$, 
respectively. By substituting
$$F_1 = \frac{1}{\sqd r} P_1 \, , ~~~F_2 = P_2 \, , ~~~G_1 = Q_1 \, , 
~~~G_2 = \frac{1}{\sqd r} Q_2 \, , $$
into Eq. (\ref{DCP}), they have the form
\begin{equation}
\begin{array}{rr}
-\pr P_1 +{\cL} P_2 = i\mu_0 r Q_1 \, , ~~
&{\cD} P_2 +{\cLd} P_1 = i\mu_0 r Q_2 \, , \\
-\pr Q_2 -{\cLd} Q_1 = i\mu_0 r P_2 \, , ~~
&{\cD} Q_1 -{\cL} Q_2 = i\mu_0 r P_1 \, . \label{reDP}
\end{array}
\end{equation}

\section{Event Horizon}

Now separating variables to Eq. (\ref{reDP}) as
$$P_1 = R_1(v,r)S_1(\theta,\varphi) \, , 
~~P_2 = R_2(v,r)S_2(\theta,\varphi) \, , $$
$$Q_1 = R_2(v,r)S_1(\theta,\varphi) \, , 
~~Q_2 = R_1(v,r)S_2(\theta,\varphi) \, , $$
then we have the radial part
\begin{equation}
\pr R_1 = (\lambda - i\mu_0r) R_2 \, , ~~
{\cD} R_2 = (\lambda + i\mu_0r) R_1 \, , \label{sepa}
\end{equation}
and the angular part
\begin{equation}
{\cLd} S_1 = -\lambda S_2 \, , 
~~{\cL} S_2 = \lambda S_1 \, ,
\end{equation}
where $\lambda = \ell +1/2$ is a separation constant. Both functions 
$S_1(\theta,\varphi)$ and $S_2(\theta,\varphi)$ are, respectively, spinorial 
spherical harmonics $_sY_{\ell m}(\theta,\varphi)$ with spin-weight $s = \pm 
1/2$, satisfying \cite{GMNRS}
\begin{eqnarray}
&&\left[\pdta +\coa \pta +\frac{1}{\sda}\pdvi +\frac{2is\cta}{\sda}\pvi
\right.\nonumber \\ 
&&\left. ~~~~ - s^2\cot^2\theta +s +(\ell -s)(\ell +s +1)\right] 
{_sY}_{\ell m}(\theta,\varphi) = 0 \, .   
\end{eqnarray}

As to the thermal radiation, we should be concerned about the behavior of 
the radial part of Eq. (\ref{sepa}) near the horizon only. Because the 
Vaidya-type space-times are spherically symmetric, we introduce as a 
working ansatz the generalized tortoise coordinate transformation 
\cite{ZD} 
\begin{equation}
r_* = r +\frac{1}{2\kappa}\ln[r -r_H(v)] \, , ~~v_* = v -v_0 \, ,\label{trans}
\end{equation}
where $r_H$ is the location of the event horizon, $\kappa$ is an adjustable
parameter and is unchanged under tortoise transformation. The parameter 
$v_0$ is an arbitrary constant. From formula (\ref{trans}), we can deduce 
some useful relations for the derivatives as follows:
$$ \pr = \left[1 +\frac{1}{\tkr}\right]\spr \, ,
~~\pv = \spv -\frac{r_{H,v}}{\tkr}\spr \, . $$

Now let us consider the asymptotic behavior of $R_1, R_2$ near the event 
horizon. Under the transformation (\ref{trans}), Eq. (\ref{sepa}) can be 
reduced to the following limiting form near the event horizon 
\begin{equation}
\spr R_1 = 0 \, , 
~~~~2r_H^2\left[G(r_H) -r_{H,v}\right]\spr R_2 = 0 \, ,  \label{trra} 
\end{equation}
after being taken limits $r \rightarrow r_H(v_0)$ and $v \rightarrow v_0$. 

From Eq. (\ref{trra}), we know that $R_1(r_*) = const$ is regular on the 
event horizon. Thus the existence condition of a non-trial solution we can
have for $R_2$ is (as for $r_H \not= 0$)
\begin{equation}
2G(r_H) -2r_{H,v} = 0 \, . \label{loca}
\end{equation}
which determines the location of horizon. The event horizon equation 
(\ref{loca}) can be inferred from the null hypersurface condition, 
$g^{ij}\partial_i F\partial_j F = 0$, and $F(v,r) = 0$, namely $r = r(v)$. 
It follows that $r_H$ depends on time $v$. So the location of the event 
horizon and the shape of the black hole change with time. 

\section{Hawking Temperature}

In the preceding section, we have deduced the event horizon equation 
from the limiting form of the separated radial part of the first order 
Dirac equation. Using a similar procedure to its second order equation, 
we can derive the Hawking temperature and the thermal radiation spectrum. 
A direct calculation gives the second order radial equation 
\begin{eqnarray} 
&&2r^2\left(G\pdr +\pdvr \right)R_1 +\left[(r^2G)_{,r} +2iqQr 
+\frac{i\mu_0\lambda -\mu_0^2r}{\lambda^2 +\mu_0^2r^2} \right]\pr R_1
-(\lambda^2 +\mu_0^2r^2)R_1 = 0 \, ,\nonumber \\
&& \label{sosr+}\\
&&2r^2\left(G\pdr +\pdvr \right)R_2 +\left[3(r^2G)_{,r} +2iqQr \right]\pr R_2
+4r\pv R_2 -\frac{i\mu_0\lambda +\mu_0^2r}{\lambda^2 +\mu_0^2r^2}\left[
2r^2G\pr \right. \nonumber \\
&&~~~~~~\left. +2r^2\pv +(r^2G)_{,r} +2iqQr \right]R_2 +\left[(r^2G)_{,rr} 
+2iqQ -(\lambda^2 +\mu_0^2r^2)\right]R_2 = 0 \, . \label{sosr-}
\end{eqnarray}

Given the transformation (\ref{trans}), Eqs. (\ref{sosr+},\ref{sosr-}) have 
the following limiting forms near the event horizon $r = r_H$
\begin{eqnarray} 
&&\left[\frac{A}{2\kappa} +4G(r_H) -2r_{H,v}\right]\spdr R_1 +2\spdvr R_1 
+\left[-A +G_{,r}(r_H) +\frac{2iqQ +2G(r_H)}{r_H} \right. \nonumber\\
&&\left. ~~~~+\frac{i\mu_0\lambda -\mu_0^2r_H}{r_H^2(\lambda^2 +\mu_0^2r_H^2)}
\right]\spr R_1 =\left[\frac{A}{2\kappa} +2G(r_H)\right]\spdr R_1 
+2\spdvr R_1 = 0 \, , \label{ra+}\\
&&\left[\frac{A}{2\kappa} +4G(r_H) -2r_{H,v}\right]\spdr R_2 +2\spdvr R_2 
+\left\{-A +3G_{,r}(r_H) +\frac{2iqQ +6G(r_H) -4r_{H,v}}{r_H} \right. 
\nonumber\\
&&\left. ~~~~-\frac{i\mu_0\lambda +\mu_0^2r_H}{r_H^2(\lambda^2 +\mu_0^2r_H^2)}
\left[2G(r_H) -2r_{H,v}\right]\right\}\spr R_2 
=\left[\frac{A}{2\kappa} +2G(r_H)\right]\spdr R_2 +2\spdvr R_2 \nonumber\\
&&~~~~+\left[-A +3G_{,r}(r_H) +\frac{2iqQ +2G(r_H)}{r_H}\right]\spr R_2 = 0 
\, . \label{ra-}
\end{eqnarray}
where we have used relations $2G(r_H) = 2r_{H,v}$ and $\spr R_1 = 0$.

With the aid of the event horizon equation (\ref{loca}), namely, $2G(r_H) 
= 2r_{H,v}$, we know that the coefficient $A$ is an infinite limit of 
$0 \over 0$ type. By use of the L' H\^{o}spital rule, we get the following 
result
\begin{equation}
A = \lim_{r \rightarrow r_H(v_0)}\frac{2(G -r_{H,v})}{r -r_H} 
= 2G_{,r}(r_H) \, .
\end{equation}

Now let us select the adjustable parameter $\kappa$ in Eqs. (\ref{ra+},
\ref{ra-}) such that
\begin{equation}
\frac{A}{2\kappa} +2G(r_H) = \frac{G_{,r}(r_H)}{\kappa} +2r_{H,v}
\equiv 1 \, ,
\end{equation}
which means the temperature of the horizon is
\begin{equation}
\kappa =\frac{G_{,r}(r_H)}{1-2G(r_H)}= \frac{G_{,r}(r_H)}{1-2r_{H,v}} \, . 
\label{temp}
\end{equation}
Such a parameter adjustment can make Eqs. (\ref{ra+},\ref{ra-}) reduce to
\begin{equation}
\spdr R_1 +2\spdvr R_1 = 0 \, ,  \label{wr1}
\end{equation}
and
\begin{equation}
\begin{array}{ll}
&\spdr R_2 +2\spdvr R_2 +\left[G_{,r}(r_H) +\frac{2iqQ +2G(r_H)}{r_H}
\right]\spr R_2 \\
&=\spdr R_2 +2\spdvr R_2 +2\left(C +i\omega_0\right)\spr R_2 = 0 \, .  
\label{wr2} 
\end{array}
\end{equation}
where $\omega_0, C$ will be regarded as finite real constants,
$$\omega_0 = \frac{qQ}{r_H} \, , 
~~2C = G_{,r}(r_H) +\frac{2r_{H,v}}{r_H} \, .$$ 
Eqs. (\ref{wr1},\ref{wr2}) are standard wave equations near the horizon. 

\section{Thermal Radiation Spectrum}

Combining Eq. (\ref{wr1}) with $\spr R_1 = 0$, we know that $R_1$ is a 
constant near the horizon. The solution $R_1 = R_{10}e^{-i\omega v_*}$ 
means that Hawking radiation does not exist for $R_1$. 

Now separating variables to Eq. (\ref{wr2}) as 
$$R_2 = R_2(r_*)e^{-i\omega v_*}$$ 
and substituting this into equation (\ref{wr2}), one gets
\begin{equation}
 R_2^{\prime\prime} = 2[i(\omega -\omega_0) -C] R_2^{\prime} \, , 
\end{equation}
The solution is
\begin{equation}
 R_2 =R_{21} e^{2[i(\omega -omega_0) -C]r_*} +R_{20} \, . 
\end{equation}

The ingoing wave and the outgoing wave to Eq. (\ref{wr2}) are 
\begin{equation}
\begin{array}{ll}
&R_2^{\rm in} = e^{-i\omega v_*} \, ,\\
&R_2^{\rm out} = e^{-i\omega v_*} 
e^{2[i(\omega -\omega_0) -C]r_*} \, ,~~~~~~~ (r > r_H) \, . 
\end{array}
\end{equation}

Near the event horizon, we have 
$$r_* \sim \frac{1}{2\kappa}\ln (r - r_H) \, .$$
Clearly, the outgoing wave $R_2^{\rm out}(r > r_H)$ is not analytic at 
the event horizon $r = r_H$, but can be analytically extended from the 
outside of the hole into the inside of the hole through the lower complex 
$r$-plane
$$ (r -r_H) \rightarrow (r_H -r)e^{-i\pi}$$
to
\begin{equation}
\tilde{R}_2^{\rm out} = e^{-i\omega v_* }e^{2[i(\omega -\omega_0) -C]r_*}
e^{i\pi C/\kappa}e^{\pi(\omega -\omega_0)/\kappa} \, ,~~~~~~(r < r_H) \, . 
\end{equation}

So the relative scattering probability of the outgoing wave at 
the horizon is easily obtained
\begin{equation}
\left|\frac{R_2^{\rm out}}{\tilde{R}_2^{\rm out}}\right|^2
= e^{-2\pi(\omega -\omega_0)/\kappa} \, . 
\end{equation}

According to the method suggested by Damour and Ruffini \cite{DR} and 
developed by Sannan \cite{San}, the thermal radiation Fermionic spectrum 
of Dirac particles from the event horizon of the hole is given by
\begin{equation} 
\langle {\cal N}_{\omega} \rangle 
= \frac{1}{e^{(\omega -\omega_0)/T_H } + 1} \, , \label{sptr}
\end{equation} 
with the Hawking temperature being 
$$ T_H = \frac{\kappa}{2\pi} \, ,$$
whose obvious expression is
\begin{equation}
T_H = \frac{1}{4\pi r_H} \cdot \frac{M r_H -Q^2  
-\Lambda r_H^4/3 }{M r_H -Q^2/2 -\Lambda r_H^4/6 } \, .
\end{equation}
It follows that the temperature depends on the time, because it is determined 
by the surface gravity $\kappa$, a function of $v$. The temperature
is consistent with that derived from the investigation of the thermal
radiation of Klein-Gordon particles \cite{LMZ}.
 
\section{Conclusions}

Equations (\ref{loca}) and (\ref{temp}) give the location and the temperature 
of event horizon, which depend on the advanced time $v$. They are just the 
same as that obtained in the discussion on thermal radiation of Klein-Gordon 
particles in the same space-time. Eq. (\ref{sptr}) shows the thermal radiation 
spectrum of Dirac particles in a charged Vaidya black hole with a cosmological
constant $\Lambda$.

In conclusion, we have studied the Hawking radiation of Dirac particles in 
a black hole whose mass and electric charge change with time. Our results 
are consistent with others. In this paper, we have dealt with the asymptotic 
behavior of the separated Dirac equation near the event horizon, not only 
its first order form but also its second order form. We find that the limiting 
form of its first order form puts very strong restrict on the Hawking
radiation, that is, not all components of Dirac spinors but $P_2, Q_1$ 
display the property of thermal radiation. The asymmetry of Hawking radiation 
with respect to the four-component Dirac spinors probaly originate from the 
asymmetry of space-times in the advanced Eddington-Finkelstein coordinate. 
This point has not been revealed previously. 

In addition, our analysis demonstrates that except the Coulomb energy 
$\omega_0$, there was no new quantum effect in a Vaidya-type space-time 
as declared by Li \cite{LMZ} . This conclusion holds true in any 
spherically symmetric black hole whether it is static or non-static.
 
\vskip 0.5cm
\noindent 
{\bf Acknowledgment}

S.Q. Wu is indebted to Dr. Jeff at Motomola Company for his long-term 
helps. This work was supported in part by the NSFC in China.


\end{document}